\documentclass[twocolumn,showpacs,preprintnumbers,prl]{revtex4}
\usepackage{amssymb}
\usepackage{citesort}

\begin{document}

\preprint{BNL-NT-03/16}
\preprint{RBRC-328}
\title{The Parton Structure of the Nucleon and Precision Determination of
the Weinberg Angle in Neutrino Scattering}
\author{Stefan Kretzer}
\affiliation{Physics Department, Brookhaven National Laboratory,
Upton, New York 11973, USA, and\\
RIKEN-BNL Research Center, Brookhaven National Laboratory,
Upton, New York 11973, USA}
\author{Fredrick Olness}
\affiliation{Department of Physics, Southern Methodist University,
Dallas, Texas 75275 USA}
\author{Jon Pumplin}
\affiliation{Department of Physics and Astronomy,
Michigan State University, East Lansing, MI 48824 USA}
\author{Mary Hall Reno}
\affiliation{Department of Physics and Astronomy,
University of Iowa, Iowa City, Iowa 52242 USA}
\author{Daniel Stump}
\affiliation{Department of Physics and Astronomy,
Michigan State University, East Lansing, MI 48824 USA}
\author{Wu-Ki Tung}
\affiliation{Department of Physics and Astronomy,
Michigan State University, East Lansing, MI 48824 USA}

\begin{abstract}
A recently completed next-to-leading-order program to calculate neutrino cross
sections, including power-suppressed mass correction terms, has been applied
to evaluate the Paschos-Wolfenstein relation, in order to quantitatively
assess the validity and significance of the NuTeV anomaly. In particular, we
study the shift of $\sin^2 \theta_{\mathrm{W}}$ obtained in calculations with
a new generation of PDF sets that allow $s(x)\neq \bar{s}(x)$, enabled by
recent neutrino dimuon data from CCFR and NuTeV, as compared to the previous
$s = \bar{s}$ parton distribution functions like CTEQ6M. The extracted value
of $\sin^2 \theta_{\mathrm{W}}$ is closely correlated with the strangeness
asymmetry momentum integral $\int_{0}^{1}x[s(x)-\bar{s}(x)] dx$. We also
consider isospin violating effects that have recently been explored by the
MRST group.  The results of our study suggest that the new dimuon data, the
Weinberg angle measurement, and other data sets used in global QCD parton
structure analysis can all be consistent within the Standard Model.
\end{abstract}

\maketitle

\paragraph{Introduction:}

An important open question in particle physics in recent years has been the
significance of the \textquotedblleft NuTeV anomaly\textquotedblright ---a
$3\,\sigma $ deviation of the measurement of
$\sin ^{2}\theta _{\mathrm{W}}$ ($0.2277\pm 0.0013\pm 0.0009$)
reported in Ref.~\cite{nuanom}, from the
world average of other measurements \cite{lepew}
($0.2227\pm 0.0004$).
Possible sources of the NuTeV anomaly, both within and beyond the
standard model, have been examined in \cite{forte}.
No consistent picture has yet emerged
in spite of extensive literature
\cite{barzo,gambino,moch,isospin,kulagin} on this subject.
The measurement in Ref.~\cite{nuanom} was based on a
correlated fit to the ratios of charged
and neutral current (CC \& NC) interactions in sign-selected neutrino and
anti-neutrino scattering events on a (primarily) iron target at Fermilab.
This procedure is closely related to measuring the Paschos-Wolfenstein ratio
\cite{pw}, which provides the theoretical underpinning of the analysis.
Specifically, the Paschos-Wolfenstein ratio $R^{-}$ is related to the
Weinberg angle $\theta_{W}$ by
\begin{eqnarray}
\label{eq:pw}
R^{-} & \equiv & \frac
{\sigma _{\mathrm{NC}}^{\nu }-\sigma _{\mathrm{NC}}^{\bar{\nu}}}
{\sigma _{\mathrm{CC}}^{\nu }-\sigma _{\mathrm{CC}}^{\bar{\nu}}}
\\ \nonumber
&\simeq& \frac{1}{2}-\sin ^{2}\theta_{\mathrm{W}}
+\delta R_{A}^{-}\ +\delta R_{QCD}^{-}+\delta R_{EW}^{-}
\end{eqnarray}
where the three correction terms are due to the non-isoscalarity of the
target ($\delta R_{A}^{-}$),
next-to-leading-order (NLO)
and nonperturbative
QCD effects ($\delta R_{QCD}^{-}$),
and higher-order electroweak effects ($\delta R_{EW}^{-}$).
Since $R^{-}$ is a ratio of differences of cross sections,
the correction terms are expected to be rather small.
But at the accuracy required to test the consistency
of the Standard Model, all the corrections need to be quantified
as precisely as possible---similar to some previous combined
perturbative/nonperturbative
re-analyses \cite{hetjets,cano} of challenges in QCD.

In this paper, we focus on QCD corrections, which are generally
recognized \cite{barzo,gambino,moch,isospin,kulagin}
to be the least well known.
Let us write
\begin{equation}
\delta R_{QCD}^{-}
=\delta R_{s}^{-}+\delta R_{I}^{-}+\delta R_{NLO}^{-}
\label{eq:QCD}
\end{equation}
where the three terms on the right-hand side are due to possible strangeness
asymmetry ($s^-=s-\bar{s}\neq 0$) and isospin violation ($u_{p,n}\neq
d_{n,p}$) effects in the parton structure of the nucleon,
and NLO ($O(\alpha_{s})$) corrections, respectively\footnote{%
Strictly speaking, there are also heavy-quark asymmetry terms such as $\delta
R_{c}^{-}\propto c-\bar{c}$ from intrinsic charm states. These must be
numerically negligible, and are certainly experimentally unknown.}. The
original NuTeV analysis was carried out at LO in QCD and assumed $\delta
R_{s}^{-}=0=\delta R_{I}^{-}$.
Our analysis is based 
the recent NLO calculation of \cite{kr03}, together with new
parton analyses that explicitly allow strangeness asymmetry ($\delta
R_{s}^{-}\neq 0$) \cite{DimuonFitting} and isospin violation ($\delta
R_{I}^{-}\neq 0$) \cite{mrst}. The actual calculation is carried out at the
cross section level, i.e., using the first line of Eq.~(\ref{eq:pw}) rather 
than using the schematic linearized form given in 
the second line of Eq.~(\ref{eq:pw}) and Eq.~(\ref{eq:QCD}). Our results provide
more realistic estimates of the sizes and uncertainties of the QCD
corrections, and a new look at the significance of the
``anomaly.''  (Cf.~also a recent re-evaluation of the electroweak correction to
the calculation of $R^-$ \cite{Hollik}.)

\paragraph{NLO Calculation:}

At sufficiently high neutrino energy,
the total neutrino cross section
\begin{equation}
\sigma ^{\nu }\equiv \sigma ^{\nu N\rightarrow lX}
=\int d^{3}p_{l}\ \frac
{d^{3}\sigma ^{\nu N\rightarrow lX}}{d^{3}p_{l}}
\label{eq:sigtot}
\end{equation}
can be calculated in QCD perturbation theory---in contrast to
charged lepton scattering, where the massless photon propagator
leads to dominance of nonperturbative photoproduction
events over deep inelastic scattering. 
The differential cross section in Eq.~(\ref{eq:sigtot})
factorizes into a sum of convolutions of parton distribution
functions and partonic cross sections
\begin{equation}
d^{3}\sigma ^{\nu N\rightarrow lX}
=\sum_{f=q,g}f\otimes d^{3}\sigma ^{\nu f\rightarrow lX}\ \ .
\label{eq:factorem}
\end{equation}

This calculation has been performed at NLO accuracy in
Ref.~\cite{kr03}.
The analysis included target and charm mass effects.
These corrections are needed to obtain reliable results because
there are non-negligible contributions from low $Q$ values to the
integral in Eq.~(\ref{eq:sigtot})---e.g., about 5\% from $Q^{2}<1\ \mathrm{%
GeV}^{2}$ for $\sigma _{\mathrm{CC}}^{\bar{\nu}}$ and from $Q^{2}<2\ \mathrm{%
GeV}^{2}$ for $\sigma _{\mathrm{CC}}^{\nu }$.
(For NC neutrino events, it
is not possible to exclude the low-$Q$ region by experimental kinematic cuts.)
Other corrections included are the non-isoscalarity of the target
material (iron), i.e., $\delta R_{A}^{-}$ in (\ref{eq:pw});
energy averaging over the neutrino and anti-neutrino flux spectra;
and cuts in hadronic energy
($20\ \mathrm{GeV}<yE_{\nu }<180\ \mathrm{GeV}$ for lepton inelasticity $y$)
as used in the experimental analysis \cite{nuanom}.

Ref.~\cite{kr03} used previously available parton distributions
\cite{ctq6,grv98}, all of which assume isospin symmetry and
$s = \bar{s}$ symmetry within the nucleon.
The study confirmed the smallness\footnote{Note, however that
NLO effects may be more important for the analysis
\cite{nuanom} that does not measure $R^-$ directly.
}
of the higher order corrections to $R^{-}$ in general.
(The same conclusion is reached by the NLO and NNLO
moment analyses of \cite{forte,moch,dobel}.)
It was also shown that the 
non-monochromatic neutrino and anti-neutrino beams,
with different profiles,
and typical cuts in the hadronic event energy
do not alter $\delta R_{NLO}^{-}$ substantially.
In the next two sections, we will examine shifts of the NLO
calculation due to recent advances in global QCD analysis
of parton distributions that allow
strangeness asymmetry and isospin violation.

In principle, the parton distribution functions in Eq.~(\ref{eq:factorem})
should be those of nuclear targets. Our calculation is done as an incoherent
sum of contributions from parton densities of unbound nucleons. This
approximation is reasonable in that we only calculate relative shifts
between $[S^-]=0$ and $[S^-] \neq 0$ PDFs (where $[S^-]$ is defined in 
Eq.~(\ref{eq:mom2});
similarly for isospin. In fact,
experimental information on nuclear PDFs is relatively scarce, and nuclear
PDFs only account for leading twist 2 ($\tau = 2$) effects. Higher twists,
whether they relate to nuclear modifications or not, are
generally difficult to handle consistently. By limiting ourselves to $\tau =
2$, our error estimates may be underestimates.

\paragraph{Strangeness Asymmetry:}

Because the strange quark mass $m_{s}$ is comparable
to $\Lambda _{\mathrm{QCD}}$,
the strange quark PDF is a nonperturbative component
of the nucleon bound state.
Except for the strangeness number sum rule,
\begin{equation}
\int \left[s(x)-\overline{s}(x)\right] dx = 0 \, ,
\label{eq:strnumsumrule}
\end{equation}
there is no fundamental or approximate symmetry that
relates the strange quark PDF $s(x)$ to the antiquark
PDF ${\bar{s}(x)}$.
Limits on $s^- \equiv s(x)-{\bar{s}(x)}$
can, therefore, only be derived from data
(or perhaps eventually from a lattice QCD calculation).
Until recently, $s^-$ has been largely unknown
and usually assumed to vanish.
However, the recently published CCFR-NuTeV
data on dimuon cross sections in $\nu N$ and
$\bar{\nu}N$ scattering yield a direct handle
on $s(x)$ and ${\bar{s}(x)}$,
and hence on $s^-$ \cite{DimuonFitting}, because the
dimuon data reflect semileptonic decays of the charm quark
in $W^{+}s\rightarrow c$ and
$W^{-}{\bar{s}}\rightarrow {\bar{c}}$ events.

An asymmetric strange sea in the nucleon ($s^-\neq 0$)
contributes to a correction term to $R^{-}$ at LO \cite{forte}.
If the scale dependence of the parton distributions is neglected, i.e.
$f(x,Q) \simeq f(x)$, and in the approximation of
overlooking experimental cuts, the
total cross section in Eq.~(\ref{eq:sigtot}) is sensitive to the second Mellin
moment integrals
$\int dx \, x \, f(x)$ of the PDFs \cite{forte,moch}.
Making the further approximation of an
isoscalar target, and in the limit of a negligible charm quark mass,
a strange sea asymmetry contributes at LO as
\begin{equation}
\delta R_{s}^{-} \simeq
- \left(
\frac{1}{2}-\frac{7}{6}\sin^{2}\theta_{\mathrm{W}}
\right)
\frac{[S^{-}]}{[Q^{-}]},
\label{eq:pws}
\end{equation}
where the strangeness asymmetry is quantified by
\begin{equation}
\left[ S^{-}\right] \equiv
\int x \left[ s(x)-{\bar{s}}(x)\right] dx \; ;
\label{eq:mom2}
\end{equation}
and $[Q^{-}]= \int x [q(x)-{\bar{q}}(x)] dx$
with $q(x)=(u(x)+d(x))/2$ represents the isoscalar up and
down quark combination.

By including the dimuon data, and by exploring the full
allowed parameter space in a global QCD analysis,
Ref.~\cite{DimuonFitting} presents a general
picture of the strangeness sector of nucleon structure.
The strong interplay between the existing experimental constraints
and the global theoretical constraints,
especially the sum rule (\ref{eq:strnumsumrule}),
places useful limits on acceptable values
of the strangeness asymmetry momentum integral $[S^{-}]$.
The limit quoted in \cite{DimuonFitting} is
$-0.001 < \left[ S^{-} \right] < +0.004$.
A large negative $[S^{-}]$ is strongly disfavored
by both dimuon and other inclusive data.
The strict sum rule (\ref{eq:strnumsumrule}) implies that
a non-zero $s^-(x)$ function must change sign at least once.
Studies in \cite{DimuonFitting} demonstrate that
the exact value of $[S^{-}]$ is a volatile quantity.
The best fit \textquotedblleft B\textquotedblright\
is a solution where negative $s^-(x)$ at low $x$
is compensated by positive $s^-(x)$ at large $x$;
this leads to positivity of the second moment integral in
Eq.~(\ref{eq:mom2}).
The same trend had previously been observed in a fit
to inclusive neutrino scattering \cite{barzo}.
Also, this behavior was anticipated by a dynamical
model \cite{broma} based on baryon-meson fluctuations of
the nucleon light-cone wave function.\footnote{%
For some more recent model discussions, cf.~e.g.~\cite{mcm}.}

We quantify the impact of the PDFs of Ref.~\cite{DimuonFitting}
on the Paschos-Wolfenstein relation in Eq.~(\ref{eq:pw})
by employing the NLO neutrino cross section calculations
of Ref.~\cite{kr03}.
The PDF sets A,B,C of Ref.~\cite{DimuonFitting}
represent good fits within the allowed parton parameter space. They all
have $s(x)\neq {\bar{s}}(x)$, and $[S^-]>0$. In our calculations, we employ
these PDFs consistently;
i.e., we use the full set of PDFs,
not just their strange quark distributions.

\begin{table}[t]
\begin{tabular}{|c|c|c|c|c|c|}
\hline
\quad fit \quad & \quad $\left[S^-\right] \times 100$ \quad
& \quad $\chi^2_{\mathrm{dimuon}}$ \quad
& \qquad $\chi^2_{\mathrm{inclusive I}}$ \qquad
& \qquad $\delta R^-_s$ \qquad
&  \\ \hline
$\mathrm{B}^+$ & 0.540 & 1.30 & 0.98 & -0.0065 &  \\ \hline
A & 0.312 & 1.02 & 0.97 & -0.0037 &  \\ \hline
B & 0.160 & \emph{1.00} & \emph{1.00} & -0.0019 &  \\ \hline
C & 0.103 & 1.01 & 1.03 & -0.0012 &  \\ \hline
$\mathrm{B}^-$ & -0.177 & 1.26 & 1.09 & 0.0023 &  \\ \hline
\end{tabular}
\caption{Shifts in $R^-$, calculated with PDF sets of
Ref.~\protect\cite{DimuonFitting}
(with non-zero $[S^-]$) compared to the value with the CTEQ6M set ($%
[S^-]=0$), are given in the last column. The quality of these new fits is
gauged by the relative $\protect\chi^2$ values (normalized to that of the
reference set ``B'') for the dimuon data set \protect\cite{dimuon} and for
the subset of global data set which have some sensitivity to $s^-(x)$
(labeled ``inclusive I''). See \protect\cite{DimuonFitting}
for details. }
\label{tab:pdfs}
\end{table}

The shift in $R^-$ due to strangeness asymmetry, $\delta R^-_s$, is obtained
as the difference:
\begin{equation}
\delta R^-_s\equiv R^-_{\{\mathrm{A,B,C,B^+,B^-}\}} - R^-_{\mathrm{CTEQ6}}\ .
\label{eq:deltar}
\end{equation}
These are given in the last column of Table \ref{tab:pdfs},
along with a summary of the underlying PDFs.
We show not only the preferred fit values for the sets
A,\,B,\,C but also results for fits $\mathrm{B}^\pm$ that were
obtained by using the Lagrange multiplier method to push the limits
of the allowed $[S^-]$ value in both directions somewhat beyond the preferred
range as described in \cite{DimuonFitting}.
The quality of the fits is indicated by the relative $\chi^2$
values, which are normalized to the reference solution ``B''.
Thus, the values in row B are 1.0 (italized) by definition.
The three preferred sets A,\,B,\,C are comparable in quality;
the extreme sets B$^{+}$ and B$^{-}$ are clearly disfavored.

For a given value of ``measured'' $R^-$, a shift of the theoretical prediction,
such as $\delta R^-_s$, leads to a shift in the extracted
$\sin^2\theta_{\mathrm{W}}$ value according to (cf.~Eq.(\ref{eq:pw})):
\begin{equation}
\delta (\sin^{2}\theta_{W}) = \delta R^-_s \ .
\end{equation}
The results of our calculation (Table 1), along with the range
$-0.001 < [S^-] < 0.004$ of
Ref.~\cite{DimuonFitting}, which is based on more extensive studies than 
just the fits shown in Table 1, lead us to estimate the range of 
$\delta R^-_s$, hence
$ \delta (\sin^{2}\theta_{W}) $, to be 
$-0.005 <  \delta (\sin^{2}\theta_{W}) < +0.001$.

We find that the shift in $R^-$, calculated as
an average
over $\nu$ and $\overline{\nu}$ energies according to their
flux spectra, is relatively insensitive to the incident
neutrino energy.
The values of $\delta R^{-}_s$ in Table \ref{tab:pdfs} are
also approximately unchanged when the cut on $yE_\nu$
is eliminated.
These findings suggest that the incorporation of other
detector effects \cite{recipe,moch}, which make the analysis
in Ref.~\cite{nuanom} somewhat more involved than a direct
measurement of $R^{-}$,
will not significantly impact the importance of the $[S^-]$
contribution to $\sin^{2}\theta_{\mathrm{W}}$\footnote{%
To estimate the size of detector-dependent effects,
we have calculated $\delta \sin^2 \theta_{\mathrm{W}}$
using the prescription of \cite{recipe},
summarized in the functional
$\int\, F[\sin^2\theta_{\mathrm{W}},s-{\bar s};x] dx$.
The shifts are within 30\% of those presented above.}.

The shift in $\sin^{2}\theta_{W}$
corresponding to the central fit B bridges a substantial part of the
original 
$3\,\sigma$ discrepancy between the NuTeV result and the
world average of other measurements of $\sin ^{2}\theta_{\mathrm{W}}$.
For PDF sets with a shift toward the negative end, such as $-0.004$,
the discrepancy is reduced to less than 
$1\,\sigma $.  On the other hand, for PDF sets with a shift toward the
positive end, such as $+0.001$, the discrepancy remains.

More input on $s^-(x)=s(x)-{\bar{s}}(x)$ would, of course, be helpful in pinning
down the contribution of strangeness asymmetry to $\delta R^{-}$.
Measurements of associated production of charmed jets and $W^{\pm }$ bosons
at the Tevatron, at RHIC, or at the future LHC, would increase our knowledge
of $s(x)$ and $\bar{s}(x)$ (cf.~\cite{wplusc}).
It will help that the \textquotedblleft
valence\textquotedblright\ density $ s^-(x)$ is more easily accessible than
the predominantly singlet $s(x)+{\bar{s}}(x)$,
which is concentrated at small $x$; however, the low expected statistics
will make this measurement extremely challenging.
In principle it seems also
feasible to study $s(x)-{\bar{s}}(x)$ on the lattice \cite{lattice}.
Unfortunately, the most relevant moment $%
[S^{-}]$ does not correspond to a local operator and cannot be calculated on
the lattice.

\paragraph{Possible Isospin Violation:}

Isospin symmetry holds to a good approximation in low energy hadron
spectroscopy and scattering, but it is not an exact symmetry. The level of
accuracy of the usual assumption of isospin symmetry at the parton level,
e.g.\ $u_{p}=d_{n}$ and $d_{p}=u_{n}$, is largely unknown. Isospin
symmetry violation effects at the parton level contribute a shift of the
P-W ratio $R^{-}$ by
\begin{equation}\label{eq:ispin}
\delta R_{I}^{-} \simeq - \left(
\frac{1}{2}-\frac{7}{6}\sin^{2}\theta_{\mathrm{W}}
\right)
\frac{[D^-_N-U^-_N]}{[Q^{-}]}
\end{equation}
where $N=(p+n)/2$ and, as before, $[\;]$ denotes the second Mellin moment.

There have been model studies \cite{isospin} that indicate $\delta R_{I}^{-}$
could be large enough to have an effect on the interpretation of the
NuTeV anomaly.  However, it is preferable to quantify the allowed
range of uncertainty of this effect directly and by model-independent global
analysis of the differences.  Unfortunately, there are few experimental
constraints on these small differences.

Nonetheless, the MRST collaboration \cite{mrst} have recently made a first
attempt to separate proton and neutron PDFs where isospin for the valence
quarks is broken by a function with a single parameter $\kappa $.
Within physically reasonable limits, they find the overall $\chi^2$ of the
global fit to be rather insensitive to $\kappa$.
By Eq.~(\ref{eq:ispin}), the determination of $\sin^{2}\theta_{W}$ via the
measurement of $R^-$ is thus subject to a non-negligible uncertainty due
to isospin violation.

To make this point more concrete, we have applied the candidate
PDFs from \cite{mrst} to our NLO calculation, in the same spirit as the
study of strangeness asymmetry discussed above.  We find that
the range of allowed $\kappa$ parameter given in \cite{mrst}, 
$-0.7<\kappa <0.7$, implies
\begin{equation}
-0.007\lesssim \delta R_{I}^{-}\lesssim 0.007 \ ,
\label{eq:mrst}
\end{equation}
and the best fit value of $\kappa =-0.2$ corresponds
to a shift of $\delta R_{I}^{-}=-0.0022$.
A one-parameter functional form may not be
general enough to pin down the \textit{true} isospin
violations of the parton structure.
Nevertheless, the large range of $\delta R^{-}_I$ in
Eq.~(\ref{eq:mrst}) indicates that a reasonable theoretical uncertainty due
to isospin violation needs to be assigned to
the determination of $\sin^{2}\theta_{\mathrm{W}}$.

\paragraph{Conclusion:}

The uncertainties in the parton structure of the nucleon that
relate to $R^{-}$ will not decrease substantially any time soon.
The uncertainties in the theory that relates $R^{-}$ to
$\sin ^{2}\theta_{\mathrm{W}}$ are substantial on the scale of
precision of the high statistics NuTeV data \cite{nuanom}.
Within their bounds, the results of this study suggest that
the new dimuon data, the Weinberg angle measurement,
and other global data sets used in QCD parton structure
analysis can all be consistent within the standard model of
particle physics.

\paragraph{Acknowledgments:}

We thank members of NuTeV collaboration, particularly K.\ McFarland, for
interesting discussions; P.\ Gambino for discussions and useful comments; and
R.\ Thorne for discussions and for providing grids of the PDFs in \cite{mrst}.
This research was supported by the National Science Foundation (grant
No.~0100677), the U.S. Department of Energy (Contracts No.~DE-FG03-95ER40908
and No.~FG02-91ER40664), and by the Lightner-Sams Foundation. S.\ K. is
grateful to RIKEN, Brookhaven National Laboratory and the U.S.~Department of
Energy (contract No.~DE-AC02-98CH10886) for providing the facilities essential
for the completion of this work.

\end{document}